\documentclass[aps,prd,reprint,nofootinbib,amssymb,floatfix]{revtex4-2}
\usepackage[tbtags]{amsmath}
\usepackage[retain-unity-mantissa = false]{siunitx}
\usepackage{mathtools,bm}
\usepackage{graphicx}	            % Include figure files
\usepackage{xcolor}\definecolor{PRDblue}{HTML}{2E3092}
\usepackage[colorlinks=true, allcolors=PRDblue]{hyperref}    
\usepackage{aas_macros}
\graphicspath{{figures/}}
\newcommand{\bhk}{\hat{\bm{k}}}
\newcommand{\bhn}{\hat{\bm{n}}}
\newcommand{\ue}{{\mathrm{e}}}
\newcommand{\ui}{{\mathrm{i}}}
\newcommand{\ud}{{\mathrm{d}}}
\newcommand{\dbn}[1][I]{\ensuremath{\delta{\bm{n}}_{#1}}}
\newcommand{\snl}[1][\mathit{s}]{\operatorname{{_{#1}{\mathcal{N}_{\ell}}}}}

\begin{document}
\title{All-sky analysis of astrochronometric signals induced by gravitational waves}
\author{Sebastian Golat}
\email{s.golat19@imperial.ac.uk}
\author{Carlo R. Contaldi}
\affiliation{Blackett Laboratory, Imperial College London, SW7 2AZ, UK}

\date{\today}
\begin{abstract}
    We introduce a unified formalism to describe both timing and astrometric perturbations induced on astrophysical point sources by gravitational waves using a complex spin field on the sphere. This allows the use of spin-weighted spherical harmonics to analyse ``astrochronometric'' observables. This approach simplifies the interpretation and simulation of anisotropies induced in the observables by gravitational waves. It also allows a simplified derivation of angular cross-spectra of the observables and their relationship with generalised Hellings-Downs correlation functions. The spin-weighted formalism also allows an explicit connection between correlation components and the spin of gravitational wave polarisations and any presence of chirality. We also calculate expected signal-to-noise ratios for observables to compare the utility of timing and deflection observables.
\end{abstract}

\maketitle%
\section{Introduction}%
Direct observation of Gravitational Waves (GWs) at \SIrange{10}{100}{Hz} is now a reality with ground-based networks of interferometers \cite{2021arXiv211103634T} routinely observing the signals emitted by the merger of massive objects at cosmological distances \cite{2021arXiv211103606T}. Future space-based interferometer missions will open a new window of GW observation at lower frequencies, \SIrange{e-5}{e-1}{Hz}, where many different signals, both galactic and extragalactic in origin, are expected to be detected \cite{2019arXiv190706482B}. The direct observation of space-time distortions offers an unprecedented opportunity for testing the nature of gravity. The detailed form of the merger signal is a direct observation of the behaviour of massive systems in the strong gravity regime and the propagation of GWs over cosmological distances and times. Current observations already give stringent constraints on the speed of propagation of GWs, modifications to the standard dispersion relation, and other extensions to the theory of General Relativity (GR) \cite{2021arXiv211206861T}. Another important test is the presence of so-called non-Einsteinian polarisations of GWs, i.e. beyond the two transverse, traceless polarisations allowed in GR.

At even lower frequencies, around \SI{1e-7}{Hz}, another window into GWs has traditionally been the measurement of the correlated perturbations of timing pulses received from pulsars across the sky \cite{Hellings}. The Pulsar Timing Array (PTA) technique is well established, and efforts monitoring a set of pulsars have been ongoing for many years \cite{2017arXiv170701615H}. This technique provides useful upper bounds on stochastic GW backgrounds \cite{10.1093/mnras/stab2833,Moore2021} and, over the past year, tantalising evidence of a potential detection has emerged \cite{Arzoumanian2020}. PTA observations are sensitive to the effect of GWs passing between the pulsars and observers. The GW distorts the effective path length taken by the pulsar signal and induces an apparent frequency shift which is observed as a perturbation of the Time-of-Arrival (ToA) of the regular pulses. This observation is most sensitive to the effect of GWs at the observer's location. A significant limitation of the PTA method is that, although timing resolution is relatively good, there is only a limited number of pulsars that we can observe and only a handful of these are intrinsically stable enough to be used for the purpose of GW monitoring. The PTA method is sensitive to the polarisation of the GWs but, given the measurement itself yields a scalar quantity, only indirectly so \cite{Cornish2018}.

Apparent distance or ToA measurements are not the only direct observables in gravitational theories based on space-time metrics. All observations involving time, distance, and \text{angular} estimates are affected by the presence of space-time perturbations. As such, another effect due to GWs is the perturbation of the apparent position of distant objects. The technique is known as the astrometric method for GW detection. This technique has so far been less prominent than the PTA method because of the limited astrometric precision of current surveys. Atmospheric effects provide a limit to the quality of astrometric observations using ground-based telescopes, which are also limited in surveying speeds and coverage. However, astrometric techniques also have some important advantages. As space-based observations increase, the precision of astrometric surveys will improve by orders of magnitude along with the speed and frequency with which the sky is surveyed. Another advantage is that any distant point source constitutes a measurement point in this technique. There are vastly more point sources in the sky than well-behaved pulsars. It is also important to note that astrometry provides an observation that is inherently directional and should therefore provide a more direct measurement of any polarisation of the underlying perturbation.

A number of authors originally considered the effect of random GWs on the apparent position of celestial objects \cite{Braginsky:1989pv,1993ApJ...418..202F,Pyne:1995iy,Pyne:1995iy,Kaiser:1996wk}. Early proposals envisaged the use of radio interferometers to monitor the relative proper motion of extragalactic objects, such as quasars, in an attempt to observe the periodic proper motion that would be induced in the presence of a random background of GWs \cite{Jaffe:2004it,Mignard2012,Darling2018}. Two contributions drive the effect \cite{Kaiser:1996wk,Gwinn_1997}. The first dominant contribution is due to the metric at the observer's location and is sensitive to GWs with frequencies up to $\sim 1/T_\mathrm{obs}$ where $T_\mathrm{obs}$ is the total duration of the observations \cite{Book2011}. The second contribution arises from the integrated effect of metric perturbations along the light-path between the source and observer. For distances that are much larger than the GW wavelength, this effect will be sub-dominant \cite{Kaiser:1996wk}.

Interesting constraints on GWs can be obtained once astrometric precision reaches \si{\micro as} precision. If we consider the root-mean-square (rms) in the correlated proper motion of $N$ sources observed with angular precision $\Delta \theta$ over a time $T_\mathrm{obs}$ induced by a Stochastic GW Background (SGWB) with spectral amplitude $\varOmega_\mathrm{gw}(f)$ we have \cite{Pyne:1995iy}
\begin{equation}
    \varOmega_\mathrm{gw}(f) \leq \frac{\Delta\theta^2}{NT_\mathrm{obs}^2H_0^2}\,,
\end{equation}
where $H_0$ is the Hubble rate and $\varOmega_\mathrm{gw}$ is the amplitude of the gravitational wave background (see below). This suggests that a survey of some $10^9$ stars, over several years at angular precision of a few \si{\micro as}, would achieve upper bounds on the cosmological background of GWs comparable to limits currently imposed by LIGO--VIRGO observations \cite{2021PhRvD.104b2004A} at higher frequencies. Today, the GAIA \cite{gaia} satellite survey is already operating with a few orders of magnitude of this baseline, and similar missions in the future may go well beyond this \cite{2021PhRvD.103h4007W}.

Different approaches have been adopted in defining the formalism through which the signal of isotropic GWs is imprinted in both timing residual and astrometric observations, including their \text{induced} anisotropies. The formalism for PTAs is well understood with the signal encoded in the correlation function known as the Hellings-Downs curve \cite{Hellings}. The analysis of the full anisotropic signal, which allows angular phase information to be preserved, is also well-developed for PTAs (see e.g. \cite{2012PhRvD..86l4028B,Roebber2017,2020PhRvD.102h4039T}). A similar approach has been taken with astrometric observations \cite{Mihaylov2018,Mihaylov2020} including the introduction of a joint, all-sky analysis of anisotropies in both timing residuals and deflections \cite{Qin2019}. This work has defined the correlation functions that are analogous to the Hellings-Downs curve for astrometry, including the correlations sourced by non-Einsteinian polarisations. The calculation of angular power spectra (the harmonic domain expansion of the correlation functions) has also been introduced.

This article extends the previous work on a joint formalism for the analysis of anisotropies in timing residuals (or redshift perturbations) and astrometric observations. We dub this formalism \text{astrochronometry} to emphasise the joint aspect of the analysis. Our work fully exploits the standard formalism for all-sky analysis of polarised observables used in the field of Cosmic Microwave Background (CMB). This is achieved by introducing spin-weighted spherical harmonics to describe the vectorial observables. The advantage of this is that the connection between correlation functions in the angular domain and angular spectra in the harmonic domain become easily generalised and entirely analogous to the relationships in the CMB signal, albeit at a different spin. The introduction of this formalism also allows us to make use of the mapping and analysis infrastructure already developed for the CMB and also allows us to define all angular spectra arising from any polarisation in the most compact and intuitive way.

This \textit{paper} is organised as follows. In Section~\ref{sec:astrochrono}, we review the signal induced by GWs in both timing residuals and deflections and define cross-correlation statistics of astrochronometry. In Section~\ref{sec:spinYlm}, we review the spin-weighted spherical harmonic formalism and show how astrochronometric observables can be re-cast into the language of complex spin-1 fields on the sphere. In Section~\ref{sec:ang}, we derive compact forms for the angular spectra sourced by all Einsteinian and non-Einsteinian polarisations. Our formalism and connection to spin-weighted expansion clarify how the
spin of the GW polarisations translates into the multipoles of the timing residuals and astrometric deflections. We do this separately for monochromatic coherent signals and a stochastic background signal. We also demonstrate how all-sky realisations of the anisotropic correctly-correlated observables can be easily obtained, having defined the appropriate spin fields. In Section~\ref{sec:snr}
we derive analytical expressions for signal-to-noise ratio statistics in astrometric observables and compare them to those for timing residuals showing the two approaches are complementary. We summarise our results in Section~\ref{sec:disc}.

\section{Astrochronometry}\label{sec:astrochrono}%
In the far-field limit of any generating mechanism and in the limit where the wavelength is much smaller than the underlying curvature scale, GWs behave as free waves perturbing the background metric. The perturbation, as experienced at some coordinate position, for a general GW, can be expressed as a sum of Fourier modes
\begin{equation}\label{eq:h}
    h_{ab}(t)=\!\int\limits_{-\infty}^{+\infty}\! \ud{f}\,\ue^{-2\pi\ui ft }\!\! \int_{S^2} \!\ud\varOmega_{\bhk}\sum\limits_{P}h_P\,(f,\bhk)\,e_{ab}^P(\bhk)\,,
\end{equation}
where $f$ is the frequency of each mode, $\bhk$ is the direction of the wave front, $h_P\,(f,\,\bhk)$ are the amplitudes of each polarisation, and $e^P_{ab}(\bhk)$ are polarisation basis tensors for each mode.\footnote{Here the polarisation index $P\in\{+,\times,X,Y,S,L\}$ runs over all polarisations being considered, potentially including non-Einsteinian ones (see appendix \ref{sec:pol}).} The spatial indices $a,b$ follow the Einstein summation convention and we adopt units where $c=1$. 

This perturbation will affect any line of sight in the sky. For example, if we monitor the ToA of regular signals like in the case of PTAs, the signal will be periodically redshifted. We can measure the integrated redshift experienced by a signal coming from the line of sight $\bhn_I$ by fitting a low-order timing model to a sequence of ToA observations and subtracting it from the observations to obtain a timing residual 
\begin{equation}\label{eq:r}
    r_I(t, \bhn_I)=\int_{0}^{t} \ud t^{\prime} z_I\left(t^{\mathrlap{\prime}}, \bhn_I\right)\,,
\end{equation}
where $z_I\left(t, \bhn_I\right)$ is the time-dependent redshift of the underlying signal's frequency along a particular line of sight. The integral recovers the overall phase shift of the signal which can be interpreted as a perturbation of the arrival time in the case of discrete pulses.

Detecting GWs using astrometry relies on repeatedly observing the apparent angular positions of many objects, preferably point sources, in the sky. Each of these positions $\bhn_I$ will be periodically deflected by GW perturbations present along the line of sight between the source and observer. We can measure this astrometric deflection over time, $\dbn(t, \bhn_I)$, and unlike in the case of PTAs, it is a quantity that is directly affected by the GW perturbation.\footnote{The two components represent the projection of the apparent shift along with the orthogonal directions $\theta$ and $\phi$ in the plane tangent to the unperturbed position and transverse to the line of sight vector.} It is also a \textit{vector} and will therefore contain inherently different projections of the underlying \textit{tensor} perturbations compared to the scalar redshifting effect.

In practice, both $z_I(t)$ and $\dbn(t)$ depend not only on metric perturbations at the point of \textit{observation} (usually the Earth) but also on perturbations at the point that we are observing (the star or pulsar)\cite{Mihaylov2018}. However, if we are only interested in a distant star, i.e. the limit where the distance to the star or pulsar is much greater than the wavelengths of the GWs, the response simplifies significantly and depends only on the perturbation at Earth \cite{Mihaylov2018}.\footnote{For polarisations with longitudinal components, this limit will introduce non-physical divergence in correlation functions at angular separations $\Theta=0$ (for details see \citet{Mihaylov2018}).} 
For notational convenience, we introduce an observation vector ${\bm{h}}_{I}(t)=({r}_{I}(t),\dbn(t))^\intercal$. While in practice we will not have both timing and astrometric observations for every line of sight, it will allow us to present a unified treatment of the two effects. 

The effect of each Fourier mode in Eq.~\eqref{eq:h} on the observed vector is well-known (see e.g. \cite{Mihaylov2018}). We can summarise this in terms of response functions.\footnote{Note that we write this in terms of vector $\bhk$ rather than the apparent position of the source of GW, $\bm{q}=-\bhk$, as used by others \citet{Mihaylov2018,Mihaylov2020}.} Starting with the response for timing residuals
\begin{equation}
    R_r^{ab}(f, \bhk,\bhn)\equiv \frac{1}{2\pi\ui{f}}{R}_z^{ab}(\bhk,\bhn) =\frac{1}{4\pi\ui{f}}\!\left(\frac{\hat{n}^{a} \hat{n}^{b}}{1+\bhk\cdot \bhn}\right)\!\label{eq:Rr}\,,\!\!
\end{equation}
where ${R}_z^{ab}(\bhk,\bhn)$ is the redshift response function, the two being related by the integral in Eq.~\eqref{eq:r}. The astrometric response function is
\begin{equation}
    R_{\hat{\imath}}^{ab}(\bhk, \bhn)=\frac{1}{2}\left[\frac{\hat{n}^{a} \hat{n}^{b}}{1+\bhk\cdot\bhn} (\hat{n}_{\hat{\imath}}+\hat{k}_{\hat{\imath}})-\delta_{\hat{\imath}}^{a} \hat{n}^{b}\right]\label{eq:Rn}\,.
\end{equation}
Notice that indices $i$ and $j$ run over the timing residual $r$ and components $\theta$ and $\phi$ of the astrometric deflection. To emphasise when we are considering only astrometric components, we will be using indices with hats that are $\hat\imath$ and $\hat\jmath$ that run only over $\theta$ and $\phi$. We will sometimes also use label $z$ to denote the response of the redshift alone without the additional frequency dependence. This allows us to write the temporal Fourier transform of the response vector as
\begin{equation}
\begin{split}
    \tilde{h}_{I,i} (f)&=\int_{S^2} \!\ud\varOmega_{\bhk}\sum\limits_{P}h_P(f,\bhk)\,R_{I,i}^{ab}(f,\bhk, \bhn_I)\,e_{ab}^P(\bhk)\,,\\
    &\equiv\int_{S^2} \!\ud\varOmega_{\bhk}\sum\limits_{P}h_P(f,\bhk)\,
     R_{I,i}^{P}(f,\bm{\hat{k}})\,. 
\end{split}
\end{equation}
Searches for GW signals in both timing and astrometry data involve cross-correlations of observations since the signal itself is mean-free. We consider the cross-correlation matrix,
$\left(h_{I,i} \star h_{J,j}\right)(\tau)=\left\langle h_{I,i}(t) h_{J,j}(t+\tau)\right\rangle$. This includes not only timing and angular correlations but also the correlation between the two.
Indices $I$ and $J$ label lines of sight to sources. 

For the case where the signal is an SGWB, we can assume that the background is statistically isotropic and that polarisation modes are independent \cite{Romano2017}
\begin{equation*}
    \langle h_P^{\phantom{\ast}}(f,\bhk)h_{P^{\prime}}^{\ast} 
    (f^{\mathrlap{\prime}},\bhk^{\mathrlap{\prime}}\,)\rangle =
    \dfrac{\delta(f-f^{\prime})}{2} \dfrac{\delta^2(\bhk, \bhk^{\mathrlap{\prime}}\,)}{4 \pi} \dfrac{\delta_{P P^{\prime}}}{g}S_{h}(f)\,,\!\!
\end{equation*}
where $S_{h}(f)$ is the one-sided strain spectral density for the individual polarisation modes and $g$ is the number of possible polarisations.\footnote{For GR polarisations $g=2$. In Section~\ref{sec:ang}, we will also consider vectorial polarisations with the same $g$ and scalar polarisations where $g=1$.} The one-sided spectral density is related to the energy density of the SGWB through
\begin{equation}
    S_{h}(f)=\frac{3 H_{0}^{2}}{2 \pi^{2}} \frac{\varOmega_{\mathrm{gw}}(f)}{f^{3}}\,.
\end{equation}
It is convenient to characterise the SGWB energy density as a power law in frequency normalised at a pivot $f_0$
\begin{equation}
    \varOmega_\text{gw}(f) = \varOmega_\text{gw}(f_0)\left(\frac{f}{f_0}\right)^\beta\,,
\end{equation}
with spectral index $\beta=0$ for scale invariant or ``cosmological'' backgrounds.
This leads to the following constraint on the cross-correlation in the Fourier domain
\begin{equation}\label{eq:h2}
    \langle\tilde{{h}}^{\phantom{\ast}}_{I,i}\!\left(f\right)\tilde{{h}}_{J,j}^\ast\!\left(f^{\prime}\right)\rangle=\frac{1}{2}\delta\left(f-f^{\prime}\right)S_{h}(f)\bar\varGamma_{IJ,ij}(f)\,,
\end{equation}
where we have introduced the overlap reduction matrix\footnote{This function has a different definition than the one used in \citet{Mihaylov2018,Mihaylov2020}, in particular it differs by a factor of $4 \pi g$. These factors are implicit in their function $T(t,t^{\prime})$, where they use $C(f)=P(f)=S_{h}(f)/4 \pi$ instead of spectral density.}
\begin{equation}    \label{eq:overlap}
    \bar\varGamma_{IJ,ij}(f)=\frac{1}{4 \pi g}\int_{S^2} \ud\varOmega_{\bhk}
    \sum\limits_{P}R_{I,i}^{P}(f,{\bhk})
    {R\mathrlap{_{J,j}}^{P}}^\ast(f,{\bhk})\,.\!\!
\end{equation}
The overlap reduction functions will, in the case of an isotropic background, depend only on $\bhn_I\cdot\bhn_J=\cos\Theta_{IJ}$. In the following sections, we will examine the angular dependence of the signal and, to that end, separate out the frequency dependence introduced by the timing components from the angular structure as follows 
\begin{equation}\label{eq:nonf}
    \bar\varGamma_{IJ,ij}(f)=\frac{(-\ui)^{\kappa_i}(\ui)^{\kappa_j}}{(2\pi f)^{\kappa_{ij}}}\varGamma_{ij}(\Theta_{IJ})  \,,
\end{equation}
where $\kappa_i=\delta_{ir}$ and $\kappa_{ij}=\kappa_i+\kappa_j$ are exponents coming from the the fact that timing residual is an integrated redshift.
Notice that this means we are effectively considering the anisotropy of the redshift effect by separating out the timing residual contribution. 

\section{Spin-weighted spherical harmonics in astrometry}\label{sec:spinYlm}
Having introduced directional dependence in the signal, it is useful to consider its angular decomposition onto spherical harmonics. The signal contains both a scalar amplitude (redshift effect) and a vector field (deflection effect) on the sphere. The scalar contribution can be decomposed onto spherical harmonics $\mathrm{Y}_{\ell m}(\vartheta,\varphi)$, the Fourier basis of the unit sphere. For the vector contribution, the conventional approach \cite{Mihaylov2020,O'Beirne2018,Mignard2012} is to expand onto the basis of vector spherical harmonics \cite{Mignard2012}. These are defined as the gradient ($\text{G}$) and curl ($\text{C}$) modes with respect to the usual scalar spherical harmonics (see e.g. \cite{Stebbins:1996wx})
\begin{alignat}{2}
    \mathbf{Y}^\text{G}_{\ell m} &=\mathbf{Y}^{E}_{\ell m}\sqrt{\ell(\ell+1)}&&= \bm{\nabla} \mathrm{Y}_{\ell m}\,,\\
    \mathbf{Y}^\text{C}_{\ell m} &=\mathbf{Y}^{B}_{\ell m}\sqrt{\ell(\ell+1)}&&= \bhn \times \bm{\nabla} \mathrm{Y}_{\ell m}\,.
\end{alignat}
Here $\vartheta$ and $\varphi$ are spherical polar coordinates and $\bhn$ is the unit vector in the direction $(\vartheta,\varphi)$. We have also introduced $E$- and $B$-modes which are often used in analogy to electromagnetic modes. The $\bm{\nabla}$ operator is defined using covariant derivatives on the sphere with metric $\mathrm{d}s^2 = \mathrm{d}\vartheta^2 + \sin^2\vartheta \,\mathrm{d}\varphi^2$ and anti-symmetric tensor
\begin{equation}
    \epsilon^i_{\,j}=\left(\!\begin{array}{cc} 0 & \sin\vartheta\\ -\csc\vartheta & 0\end{array}\,\right)\,.
\end{equation}

A vector field $\bm{V}(\bhn )$ on the unit sphere is then decomposed as
\begin{align}\label{eq:alm1}
    a_{\ell m}^E&= a_{\ell m}^\text{G} \sqrt{\ell(\ell+1)}= \int_{S^2} \ud\varOmega_{\bhn}\frac{   {\mathbf{Y}_{\mathrlap{\ell m}}^{\text{G}}}^{\ast}(\bhn)\cdot\bm{V}(\bhn)}{\sqrt{\ell(\ell+1)}} \,,\\\label{eq:alm2}
    a_{\ell m}^B&= a_{\ell m}^\text{C} \sqrt{\ell(\ell+1)}= \int_{S^2} \ud\varOmega_{\bhn}\frac{{\mathbf{Y}_{\mathrlap{\ell m}}^{\text{C}}}^{\ast}(\bhn)\cdot \bm{V}(\bhn) }{\sqrt{\ell(\ell+1)}} \,.
\end{align}

It is convenient to relate this expansion to that of spin-weighted spherical harmonics \cite{Newman:1966ub,Goldberg}. Spin-weighted spherical harmonics are used to expand the spin-$s$ fields $_sF(\bhn)$ on the sphere. Spin-$s$ fields are defined as quantities that transform as $e^{-\ui s\psi}$ under right-handed rotations about $\bhn$ by an angle $\psi$.

Polarisation of Cosmic Microwave Background (CMB) radiation is a well-known application of spin-weighted spherical harmonics. The linear polarisation of CMB radiation arriving along a line of sight $\bhn$ is described by Stokes parameters, $Q$ and $U$ \cite{Jackson1998}. These can be packaged into complex variables $_{\pm 2}F(\bhn)\equiv (Q\pm \ui U)(\bhn)$ that transform as a spin-$s$ field on the unit sphere with $s=\pm 2$ \cite{Zaldarriaga:1996xe}. This is possible because they are quadratic in the underlying spin-$1$ electromagnetic field. Similar quantities can be defined for tensor polarisations such as the transverse-traceless modes of GW radiation. GW $Q$ and $U$ Stokes parameters transform as spin-4 fields as they are quadratic in the underlying spin-$2$ tensor field. However, GW Stokes parameters can only be observed by making direct measurements of the GWs (see e.g. \cite{Arianna&Contaldi}).
Astrochronometric measurements only depend on the \textit{projected} effect of the underlying polarisation modes. In order to avoid confusion with GW Stokes parameters, we do not introduce $Q$ and $U$ in defining Stokes parameter analogues for astrometry. However, the analogy is straightforward if we introduce deflection components $\delta{\tilde{n}}_{\theta}$ and $\delta{\tilde{n}}_{\phi}$ of the Fourier decomposed observation vector ${\bm{h}}_{I}$ defined above. 

The advantage of introducing spin-weighted fields is that raising and lowering operators, $\eth$ and $\bar{\eth}$ respectively, can be defined. These can be used to raise or lower the spin of a field to obtain spin-$0$ (scalar) quantities that are invariant under rotations of the coordinate system. The raising and lowering operators are defined through the operations
\begin{alignat}{3}
    \!\!\eth [{}_sF(\bhn)]&=-\sin^{s}\!\vartheta
    &&\!\left(\frac{\partial}{\partial\vartheta}+\frac{\ui}{\sin{\vartheta}}
    \frac{\partial}{\partial\varphi} \right)\![\csc^{s}\!{\vartheta} \,&&{}_sF(\bhn)]\,,\!\!
     \\
   \!\! \bar{\eth}[ {}_sF(\bhn)]&=-\csc^{s}\!{\vartheta}
    &&\!\left(\frac{\partial}{\partial\vartheta}-\frac{\ui}{\sin{\vartheta}}
    \frac{\partial}{\partial\varphi} \right)\![\sin^{s}\!{\vartheta} \, &&{}_sF(\bhn)]\,. \!\!
\end{alignat}
The vector field $\bm{V}(\bhn )$ can be decomposed into spin-$1$ fields $ _{\pm 1}F(\bhn)\equiv \delta{\tilde{n}}_{\theta}\pm \ui \delta{\tilde{n}}_{\phi}$ with  $\delta{\tilde{n}}_{\theta}=\delta\Tilde{\bm{n}}\cdot \hat{\bm{e}}_\vartheta$ and 
$\delta{\tilde{n}}_{\phi}= \delta\Tilde{\bm{n}}\cdot \hat{\bm{e}}_\varphi$ and where $\hat{\bm{e}}_\vartheta$ and $\hat{\bm{e}}_\varphi$ are the orthogonal unit vectors transverse to the radial direction $\bhn$. The spin-1 combinations can be lowered and raised to spin-0 quantities and expanded onto spherical harmonics as a set of spin-1 coefficients $_{\pm 1}a_{\ell m}$
\begin{align}
    \bar{\eth} (\delta{\tilde{n}}_{\theta}+\ui \delta{\tilde{n}}_{\phi})(\bhn) &= \sqrt{\ell(\ell+1)}{\,}_{+1}a_{\ell m}\mathrm{Y}_{\ell m}(\bhn) \,,\\
     \eth(\delta{\tilde{n}}_{\theta}-\ui \delta{\tilde{n}}_{\phi})(\bhn) &= \sqrt{\ell(\ell+1)}{\,}_{-1}a_{\ell m}\mathrm{Y}_{\ell m}(\bhn)\,.
\end{align}
The expansion can be inverted using
\begin{align}\label{eq:salm1}
    _{+1}a_{\ell m } &= \int_{S^2} \ud\varOmega_{\bhn}\,  \frac{\bar{\eth}\mathrm{Y}^\ast_{\ell m}(\bhn)}{\sqrt{\ell(\ell+1)}}(\delta{\tilde{n}}_{\theta}+\ui \delta{\tilde{n}}_{\phi})(\bhn)\notag\\&\equiv \int_{S^2} \ud\varOmega_{\bhn}\, {}_{+1}\!\mathrm{Y}^\ast_{\ell m}(\bhn)(\delta{\tilde{n}}_{\theta}+\ui \delta{\tilde{n}}_{\phi})(\bhn)\,,\\\label{eq:salm2}
    _{-1}a_{\ell m } &= \int_{S^2} \ud\varOmega_{\bhn}\, \frac{\eth \mathrm{Y}^\ast_{\ell m}(\bhn)}{\sqrt{\ell(\ell+1)}}(\delta{\tilde{n}}_{\theta}-\ui \delta{\tilde{n}}_{\phi})(\bhn)\notag\\&\equiv \int_{S^2} \ud\varOmega_{\bhn}\, {}_{-1}\!\mathrm{Y}^\ast_{\ell m}(\bhn)(\delta{\tilde{n}}_{\theta}-\ui \delta{\tilde{n}}_{\phi})(\bhn)\,,
\end{align}
where we have defined spin-1 spherical harmonics $_{\pm 1}\!\mathrm{Y}_{\ell m}$. 
Comparing Eqs.~\eqref{eq:alm1} and \eqref{eq:alm2} with Eqs.~\eqref{eq:salm1} and \eqref{eq:salm2}, spin-weighted modes can be combined into $E$ and $B$ modes as \footnote{The definition requires a choice of convention. We use the same choice used in the \texttt{Healpix} package \cite{2005ApJ...622..759G}.}
\begin{alignat}{2}
    a_{\ell m}^E &=  \frac{1}{2}&&(_{-1}a^{\,}_{\ell m} - {}_{1}a^{\,}_{\ell m})\,,\\ 
    a_{\ell m}^B &= \frac{\ui}{2}&&(_{-1}a^{\,}_{\ell m} + {}_{1}a^{\,}_{\ell m})\,.
\end{alignat}
Note that $_{\pm 1}\!\mathrm{Y}_{\ell m}=0$ for $\ell = 0$. Using these definitions we can directly expand the spin-$1$ field as
\begin{equation}\label{eq:aEB}
    (\delta{\tilde{n}}_{\theta} \pm \ui \delta{\tilde{n}}_{\phi})({\bhn})=\sum_{\ell m}\left(\mp a_{\ell m}^{E} - \ui a_{\ell m}^{B}\right) {}_{\pm 1}\!\mathrm{Y}_{\ell m}(\bhn)\,.
\end{equation}

The advantage of introducing spin-weighted decompositions is that existing computational frameworks developed for CMB analysis can be adapted for studies in this field. For example, the \texttt{HEALPix} package \cite{2005ApJ...622..759G} implements spherical harmonic transforms of spin-$s$ fields using spin-weighted basis functions and can be used to generate a realisation of the anisotropic sky with properly correlated timing and astrometric signals, as we show below. This formalism can also be used in the analysis of future observations. In particular, the ability to re-bin observations and to apply smoothing and other angular filters to astrochronometric data will aid the internal analysis and the cross-correlation with other observables.

The use of complex spin-$1$ fields also simplifies the interpretation of angular power spectra of the signal by analogy with CMB spectra and clarifies the link between the parity of underlying tensor polarisation and the different spin-$1$ cross-correlation spectra as we also show below. 

\section{Angular power spectra and correlation functions}\label{sec:ang}
A convenient way to study the angular behaviour of correlations in astrochronometry is to use the angular power spectra. Angular spectra are particularly useful for observations that cover large portions of the sky and can help to understand the angular dependence of different correlations. A well-known application is the prediction of CMB spectra. In astrochronometry, angular spectra are the Fourier domain conjugates of the Hellings-Downs curve \cite{Hellings} (for timing residuals) and its analogues for astrometry. Angular spectra have been previously introduced in both PTA and astrometry analysis \cite{Roebber2017,Mihaylov2020,O'Beirne2018,Qin2019}. Here, we review their application and extend the calculation to the unified framework we have adopted in this work, emphasising the relationship between different angular cross-correlations and the underlying tensor polarisations.

Following \citet{Roebber2017}, we identify the redshift \textit{scalar} field as the analogue to the intensity or temperature $T$. Here we can loosely relate $T$ to ``{timing}''. The redshift field can be expanded as
\begin{equation}\label{eq:aT}
    \tilde{z}(\bhn)=\sum_{\ell m} a^{T}_{\ell m} \mathrm{Y}^{\,}_{\ell m}({\bhn})\,, 
\end{equation}
noting that, as usual, we have separated out the frequency dependence, as discussed previously. We focus on the redshift signal for now as additional considerations are required when expanding timing residuals due to the integration involved.

We can form several angular cross-correlation spectra
\begin{equation}\label{eq:CQQ}
    C^{QQ^\prime}_{\ell}=\frac{1}{2 \ell+1} \sum\limits_{m=-\ell}^{\ell} a_{\ell m}^{Q} {a_{\mathrlap{\ell m}}^{Q^\prime}}^\ast\,,
\end{equation}
considering the full set of modes, $a_{\ell m}^{Q} \in \{a_{\ell m}^{T},a_{\ell m}^{E},a_{\ell m}^{B}\}$.
Notice that, in the most general case, the cross-spectra are complex-valued but any imaginary component is generated solely by chiral components in the underlying signals such as circularly polarised GWs (see below). This, in turn, induces specific patterns in the cross-correlation of scalar and spin-1 observables.

The spectra contain the full statistical description of anisotropies, assuming statistical isotropy and Gaussianity. In particular, a compression to angular spectra assumes there is no information in the angular phase of the modes in the sky. The spectra also have the useful property of being invariant under rotations. This is the most useful compression for the case where the underlying signal is due to a stochastic background of GWs. However, it can also be applied to the case of a signal due to a single monochromatic wave. In this case, the signal is not invariant under rotations, but it is still useful to compress to angular spectra if we are not interested in the angular phase information.

\subsection{Coherent monochromatic signals}%
We first consider the signal due to a monochromatic wave. For convenience let us choose a wave aligned with the north pole, i.e. $\bhk=-\hat{\bm{z}}$, with Einsteinian polarisation components $h_+$ and $h_\times$, and a star at angular coordinates $(\vartheta,\varphi)$ in direction $\bhn$. The response will be
\begin{equation}\label{eq:ztt}
    \tilde{z}(\bhn)=\frac{1}{2}(1+\cos{\vartheta})\left({h}_{+}\cos{2 \varphi} -{h}_{\times}\sin{2 \varphi}  \right)\,,
\end{equation} for redshift and
\begin{equation}\label{eq:dntt}
    (\delta{\tilde{n}}_{\theta} \pm \ui \delta{\tilde{n}}_{\phi})({\bhn})=\frac{1}{2}\ue^{\pm2\ui\varphi}\sin{\vartheta}\left({h}_{+}\pm\ui {h}_{\times}\right),
\end{equation}
for the deflection. For notational convenience, we introduce the normalisation of 
spin-$s$ polarisation
\begin{equation}
    \snl=\sqrt{\frac{(\ell-s)!}{(\ell+s)!}}\,,
\end{equation}
which appears as general $\ell$-scaling of both $a_{\ell m}$ and $C_{\ell}$.
Eqs.~\eqref{eq:aT}~and~\eqref{eq:aEB} can now be inverted to obtain the harmonic coefficients
\begin{align}
    a_{\ell m}^{T}&=2 \pi \snl[2]\sqrt{\frac{2 \ell+1}{4 \pi} }\left({h}_{+} \pm \ui {h}_{\times}\right)\delta_{m\pm2}\,\,,\\
    a_{\ell m}^{E}&=2\,a_{\ell m}^{T}\sqrt{\frac{(\ell-1)!}{(\ell+1)!}}
    \phantom{\ui}{\left(\delta_{m-2}+\delta_{m2}\right)}\,,\\
    a_{\ell m}^{B}&=2\,a_{\ell m}^{T}\sqrt{\frac{(\ell-1)!}{(\ell+1)!}}\,\ui{\left(\delta_{m-2}-\delta_{m2}\right)}\,.
\end{align}
Due to the spin-2 nature of the polarisation, which is manifest in Eqs.~\eqref{eq:ztt}~and~\eqref{eq:dntt}, the modes only contribute to the $m=\pm 2$ pole at each scale $\ell$. This property is not invariant under a general orientation of the coordinate system. However, the consequence of this is invariant under rotations, i.e. no monopole or dipole present in the anisotropies.
We can obtain angular power spectra 
using Eq.~\eqref{eq:CQQ} which will be proportional to $I=|{h}_{+}|^2 + |{h}_{\times}|^2$. The angular power spectrum of redshift--redshift correlations is
\begin{equation}
    C^{TT}_{\ell\ge2}= {2\pi}I\snl[2]^2.
\end{equation}
In the case of the redshift--deflection cross-correlation, there will be an additional $\ell$ dependence coming from the vector nature of one of the observables, which gives
\begin{equation}
    C^{TE}_{\ell\ge2}=\frac{{4\pi}I}{\sqrt{\ell(\ell+1)}}\snl[2]^2.\\
\end{equation}
Deflection--deflection $E$ and $B$ correlations will be equal,
\begin{equation}
    C^{EE}_{\ell\ge2}\equiv C^{BB}_{\ell\ge2}=\frac{{8\pi} I}{{\ell(\ell+1)}}\snl[2]^2, 
\end{equation}
and will have appropriate $\ell$-scaling coming from the quantities being correlated.
The parity-violating correlations will instead depend on $V=-2\operatorname{Im}\{{h}_{+}{h}_{\times}^\ast\}$ as
\begin{align}
    C^{TB}_{\ell\ge2}&=-\frac{{4\pi{\ui}}V}{\sqrt{\ell(\ell+1)}}\snl[2]^2,\\
    C^{EB}_{\ell\ge2}&=-\frac{{8\pi{\ui}}V}{{\ell(\ell+1)}}\snl[2]^2.
\end{align}

We can also repeat these steps for any additional non-Einsteinian polarisations. The six polarisation basis tensors are defined in Appendix~\ref{sec:pol}. We shall start with vector polarisations for which
\begin{equation}
    \tilde{z}(\bhn)=\frac{1}{2}\frac{\sin{2\vartheta}}{1-\cos{\vartheta}}\left({h}_{X}\cos{\varphi} +{h}_{Y}\sin{\varphi}\right)\,,
\end{equation}
and
\begin{equation}\begin{split}
     (\delta{\tilde{n}}_{\theta} \pm \ui \delta{\tilde{n}}_{\phi})({\bhn})&=\frac{1}{2}\!\left[
    (1+2 \mu)(h_{X}\cos \varphi-h_{Y}\sin \varphi)\right.\!\!\\
    &\left.\phantom{=}\pm\ui \mu(h_{X}\sin \varphi+h_{Y}\cos \varphi) 
    \right]\!\,,\!\!
\end{split}
\end{equation}
where $\mu=\cos{\vartheta}$. This time, the $\vartheta$-integral will lead to a different value for $\ell=1$ than for the other modes 
\begin{align}
    \!a_{\ell m}^{T}&= 2\pi\!\snl[1]\!\sqrt{\frac{2 \ell+1}{4 \pi}}(\ui {h}_{Y}\pm{h}_{X})\Bigl(\frac{2}{3}\delta_{\ell1}-1\Bigr)\delta_{m\pm1}\,,\!\!\\
    \!a_{\ell m}^{E}&=a_{\ell m}^{T}\sqrt{\frac{(\ell-1)!}{(\ell+1)!}}
    \phantom{\ui}{\left(\delta_{m-1}+\delta_{m1}\right)}\,,\\
    \!a_{\ell m}^{B}&= a_{\ell m}^{T} \sqrt{\frac{(\ell-1)!}{(\ell+1)!}}\,\ui{\left(\delta_{m-1}-\delta_{m1}\right)}\,.
\end{align}
Again there are only coefficients with $m=\pm1$ due to the spin-1 nature of the vectorial polarisation, which means that this time we will also have dipole correlations that will be different from the higher moments. 

The angular power spectrum redshift--redshift correlations will again be proportional to the intensity of the vectorial GW, $I=|{h}_{X}|^2 + |{h}_{Y}|^2$, giving
\begin{equation}
    C^{TT}_{\ell\ge1}= {2\pi}I\left(1-\frac{8}{9}\delta_{\ell1}\right)\snl[1]^2.
\end{equation}
The redshift cross-correlation with $E$-mode will have no factor of two this time and the dipole will be anti-correlated, which can be written as
\begin{equation}
    C^{TE}_{\ell}=\frac{{2\pi}I}{\sqrt{\ell(\ell+1)}}\left(1-\frac{10}{9}\delta_{\ell1}\right)\snl[1]^2.
\end{equation}
The astrometric $E$ and $B$ correlations will be equal, but there won't be the extra factor of four, that is
\begin{equation}
    C^{EE}_{\ell}=C^{BB}_{\ell}=\frac{{2\pi}I}{{\ell(\ell+1)}}\left(1-\frac{8}{9}\delta_{\ell1}\right)\snl[1]^2.
\end{equation}

For vectorial parity-violating modes, the Stokes parameter $V=-2\operatorname{Im}\{{h}^{\,}_{X}{h}_{Y}^\ast\}$ and angular spectra are
\begin{align}
    C^{EB}_{\ell}&=-\frac{{2\pi{\ui}}V}{{\ell(\ell+1)}}\left(1-\frac{10}{9}\delta_{\ell1}\right)\snl[1]^2, \\
    C^{TB}_{\ell}&=-\frac{{2\pi{\ui}}V}{\sqrt{\ell(\ell+1)}}\left(1-\frac{8}{9}\delta_{\ell1}\right)\snl[1]^2.
\end{align}

The scalar transverse polarisation (breathing mode) redshifts signal as
\begin{equation}
    \tilde{z}(\bhn)=\frac{1}{2}{h}_{S}(1+\cos{\vartheta})\,,
\end{equation}
and deflects apparent positions
\begin{equation}
    (\delta{\tilde{n}}_{\theta} \pm \ui \delta{\tilde{n}}_{\phi})({\bhn})=\frac{1}{2}{h}_{S}\sin{\vartheta}\,.
\end{equation}
In this case, due to the even parity of any scalar modes, there are no $B$-modes in the deflection anisotropies (see Appendix~\ref{sec:pol}), and we obtain
\begin{align}
    a_{\ell m}^{T}&= 2\pi \sqrt{\frac{2 \ell+1}{4 \pi}}h_S\left(\delta_{\ell0}+\frac{\delta_{\ell1}}{3}\right)\delta_{m0}\,,\!\!\\
    a_{\ell m}^{E}&=2a_{\ell m}^{T}\sqrt{\frac{(\ell-1)!}{(\ell+1)!}}\,\delta_{\ell1}\,.
\end{align}
For this polarisation, only monopole and dipole contributions are non-zero: 
\begin{align}
    C^{TT}_{\ell}&=\pi|h_S|^2\left(\frac{\delta_{\ell1}}{9}+\delta_{\ell0}\right)\,,
    \\
    C^{TE}_{\ell}&=\frac{2\pi|h_S|^2}{\sqrt{\ell(\ell+1)}}\frac{\delta_{\ell1}}{9}\,,
    \\
    C^{EE}_{\ell}&=\frac{4\pi|h_S|^2}{\ell(\ell+1)}\frac{\delta_{\ell1}}{9}\,.
\end{align}%
Note that the monopole in redshift is not an interesting observable as it is degenerate with the reference timing or frequency standard.

The scalar longitudinal polarisation gives redshift
\begin{equation}\label{eq:zSL}
    \tilde{z}(\bhn)=\frac{1}{2}\frac{\sqrt{2}\cos^2{\vartheta}}{1+\cos{\vartheta}}{h}_{L}\,,
\end{equation}
and astrometric deflections
\begin{equation}
    (\delta{\tilde{n}}_{\theta} \pm \ui \delta{\tilde{n}}_{\phi})({\bhn})=\frac{\sqrt{2}\sin{2\vartheta}}{1+\cos{\vartheta}}{h}_{L}\,.
\end{equation}
Longitudinal polarisation is different from the others. There is no $a_{\ell m}^{T}$ because integral of Eq.~\eqref{eq:zSL} over $\vartheta$ diverges thanks to the distant star limit. It can be removed by considering finite distances to the sources but this is beyond the scope of this work. Notwithstanding this, we can still obtain an $E$-mode contribution to deflections,
\begin{equation}
    a_{\ell m}^{E}=2\pi \sqrt{\frac{2 \ell+1}{4 \pi} \frac{(\ell-1) !}{(\ell+1) !}}h_L\sqrt{2}\left(1-\frac{2}{3}\delta_{\ell1}\right)\delta_{m0}\,,
\end{equation}
which leads to the angular power spectrum
\begin{equation}
  C^{EE}_{\ell}=\frac{{2\pi}|h_L|^2}{\ell(\ell+1)}
  \left(1-\frac{8}{9}\delta_{\ell1}\right)\,.
\end{equation}

\subsection{Statistically isotropic backgrounds}\label{sec:SGWB}%
In the case of stochastic backgrounds that are statistically isotropic, we instead expand the responses ${R}^P_{I,z}(\bhk)$ in spherical harmonics and ${R}^P_{I,\pm}(\bhk)={R}^P_{I,\theta}(\bhk)\pm\ui{R}^P_{I,\phi}(\bhk)$ in spin-1 weighted harmonics in order to obtain expansions for the overlap reduction functions.\footnote{Notice that the harmonic expansion is with respect to the line or sight $\bhn_I$, not the wave vector $\bhk$.}

Due to statistical isotropy, the correlations in the expanded modes are diagonal, that is 
\begin{equation}\label{eq:ensemble}
    \left\langle a_{\ell m}^{P,Q}(\bhk) \, 
    {a_{\mathrlap{\ell^{\prime} m^{\prime}}}^{P^{\prime}\!\!,Q^\prime}}^\ast\!\!
    (\bhk)\right\rangle=\delta_{\ell \ell^{\prime}} \delta_{m m^{\prime}}\delta_{P P^{\prime}} C_{\ell}^{QQ^\prime}\,.
\end{equation}
Notice that in defining the (frequency-independent) overlap reduction functions, we integrate over the wave vectors in Eq.~\eqref{eq:overlap}, which effectively adds an angular averaging to the ensemble average, giving
\begin{equation}    
    \begin{split}
        \varGamma_{ij}(\bhn_{I},\bhn_{J})&=\left\langle R_{I,i}^{P}({\bhk})
        R_{J,j}^{P\ast}({\bhk})\right\rangle\\
        &=\frac{1}{4 \pi g}\int_{S^2} \!\ud\varOmega_{\bhk}
        \sum\limits_{P}
        R_{I,i}^{P}({\bhk})
        R_{J,j}^{P\ast}({\bhk})\label{eq:ord}\,,
    \end{split}
\end{equation}
where $\bhn_I\cdot\bhn_J=\cos{\Theta_{IJ}}$ and 
the average is over all directions $\bhk$ and polarisations $P$ in the ensemble considered, and $g$ is the number of these polarisations.\footnote{This is equal to two for GR and vectorial polarisations and equal to one for scalar polarisations.}

\begin{figure*}[!t]
    \centering
    \includegraphics[width=0.48\textwidth]{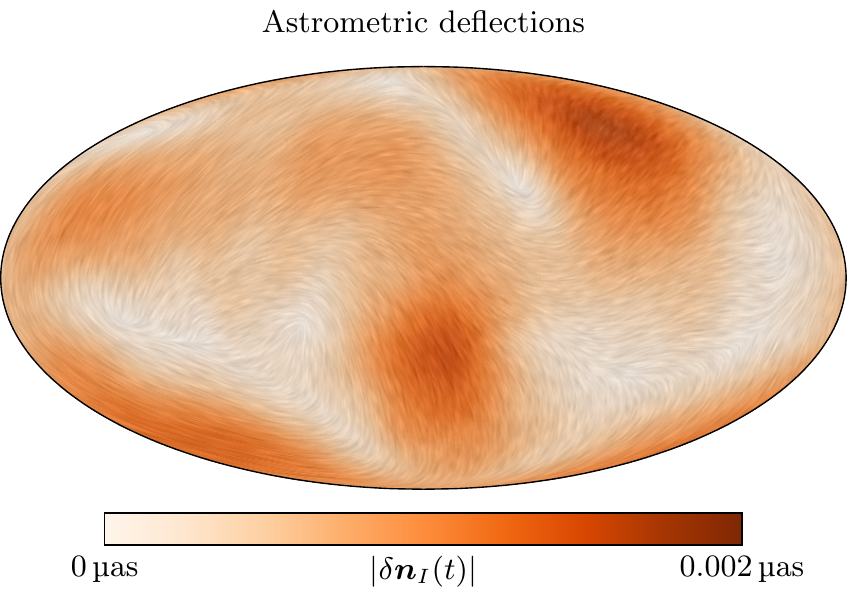}\quad
    \includegraphics[width=0.48\textwidth]{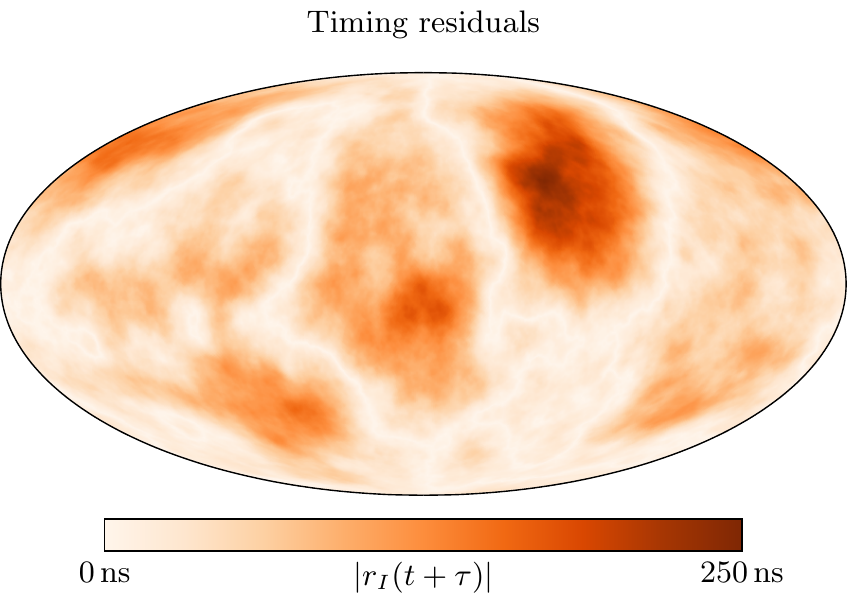}
    \caption{Realisations of astrometric deflection (at time $t$) and redshift (at time $t+\tau$) responses to an SGWB of cosmological origin (with spectral index $\beta=0$) and $\varOmega_\mathrm{gw}(f_0=\SI{50}{Hz})=\num{1e-8}$ produced using the \texttt{HEALPix} package. The time lag $\tau=\SI{43}{weeks}$ between the maps has been chosen to emphasise the cross-correlation.}
    \label{fig:gaiamap}
\end{figure*}%

The analogy with CMB spectra and correlation functions is particularly useful in this case as we can define Legendre transforms between the line of sight (pixel) and harmonic domain using \cite{Ng1999,Contaldi2016}:
\begin{subequations}
    \begin{align}
        \begin{split}
            \!\!\left\langle\!{{R}_{I,\mathrlap{z}}^P}\phantom{^{\ast}}{{R}_{J,\mathrlap{z}}^P}^{\ast}\right\rangle 
            &=\sum_{\ell} \!\frac{2 \ell\mkern -1.5mu +\mkern -1.5mu 1}{4 \pi} C_{\ell}^{TT}d^{\ell}_{00}(\mkern -1mu \Theta_{I\mkern -1.5mu J}\mkern -1.5mu)
            \\
            &=\varGamma_{zz}(\mkern -1mu \Theta_{I\mkern -1.5mu J}\mkern -1.5mu)\,,\label{eq:Czz}
        \end{split}\\
        \begin{split}
            \!\!\left\langle\! {{R}_{I,\mathrlap{z}}^P}\phantom{^{\ast}}{{R}_{J,\mathrlap{\pm}}^P}^{\ast}\right\rangle
            &=\!\sum_{\ell} \!{\frac{2 \ell\mkern -1.5mu +\mkern -1.5mu 1}{4 \pi}}\!\left(C_{\ell}^{T E}\!\mkern -1.5mu \pm\ui C_{\ell}^{T B}\right)\! d^{\ell}_{10}(\mkern -1mu \Theta_{I\mkern -1.5mu J}\mkern -1.5mu)
            \\
            &=\left(\varGamma_{z\theta}\pm\ui\varGamma_{z\phi}\right)(\mkern -1mu \Theta_{I\mkern -1.5mu J}\mkern -1.5mu) \,,
        \end{split}\\
        \begin{split}
            \!\!\left\langle\!{{R}_{I,\pm}^P\!\!\!}\phantom{^{\ast}}{{R}_{J,\mathrlap{\pm}}^P}^{\ast}\right\rangle
            &=\sum_{\ell} \!{\frac{2 \ell\mkern -1.5mu +\mkern -1.5mu 1}{4 \pi}}\!\left(C_{\ell}^{EE}\!\mkern -1.5mu+C_{\ell}^{BB}\right)\! d^{\ell}_{11}(\mkern -1mu \Theta_{I\mkern -1.5mu J}\mkern -1.5mu)\
            \\
            &=\left(\varGamma_{\theta\theta}+\varGamma_{\phi\phi}\right)(\mkern -1mu \Theta_{I\mkern -1.5mu J}\mkern -1.5mu)\,,
        \end{split}\\
        \begin{split}
            \!\!\left\langle\! {{R}_{I,\pm}^P\!\!\!}\phantom{^{\ast}}{{R}_{J,\mathrlap{\mp}}^P}^{\ast}\right\rangle
            &=\sum_{\ell} \!{\frac{2 \ell\mkern -1.5mu +\mkern -1.5mu 1}{4 \pi}}\!\left(C_{\ell}^{EE}\!\mkern -1.5mu -C_{\ell}^{BB}\!\mkern -1.5mu \pm2\ui C_{\ell}^{EB}\right)\!d^{\ell}_{1-\mkern -1.5mu 1}(\mkern -1mu \Theta_{I\mkern -1.5mu J}\mkern -1.5mu)\!\!
            \\
            &=\left(\varGamma_{\theta\theta}-\varGamma_{\phi\phi}\pm\ui\varGamma_{\theta\phi}\pm\ui\varGamma_{\phi\theta}\right)(\mkern -1mu \Theta_{I\mkern -1.5mu J}\mkern -1.5mu)\,,\!\!\label{eq:Cnn}
        \end{split}
    \end{align}
\end{subequations}%
where $d^\ell_{mm^\prime}$ are the \textit{small} Wigner rotation operators \cite{Khersonskii:1988krb}. These expressions give an intuitive understanding of how the Hellings-Downs curve, and its analogues for astrometry for any polarisation, are related to even and odd parity spectra. In turn, this gives an understanding of how the different correlations are sourced by the parity of the underlying GW polarisations which are analogous to their CMB counterparts. 

For example, using Eqs.~\eqref{eq:Czz}--\eqref{eq:Cnn}, we can easily infer which unique signatures in the overlap reduction functions 
would be produced by a chiral GW background with parity-violating modes $C_\ell^{TB}$ and $C_\ell^{EB}$.

The spectra $C_\ell^{EE}$ and $C_\ell^{BB}$ were calculated in \cite{Mihaylov2020} for tensorial, vectorial and scalar polarisations using the formalism introduced by \citet{O'Beirne2018}. In fact, all spectra can be calculated as simple scaling laws in multipole $\ell$ {\textit a priori}.

For tensorial polarisations, power spectra will be zero for multipoles lower than a quadrupole, leaving
\begin{align}
    C^{TT}_{\ell \geq 2} &= 2\pi\snl[2]^2,\\
    C^{TE}_{\ell \geq 2} &= \frac{4\pi}{\sqrt{\ell(\ell+1)}}\snl[2]^2,\\
    C^{EE}_{\ell \geq 2} &= C^{BB}_{\ell \geq 2} = \frac{8\pi}{\ell(\ell+1)}\snl[2]^2.\label{eq:CET}\
\end{align}
Notice that the angular spectra do not contain any information about the SGWB amplitude but only on the anisotropic correlation induced by the observables. The overall normalisation of the correlation patterns is provided by the spectral density $S_h(f)$ in Eq.~\eqref{eq:h2} (see Appendix~\ref{sec:appendixA}). 

It is easy to check (see Table~\ref{tab:tensor} in Appendix~\ref{sec:gamma}) that Eq.~\eqref{eq:CET} is the same as the $C_\ell$ presented in \citet{Mihaylov2020} up to a factor of two.\footnote{This factor is just a convention. In \citet{Mihaylov2020} they use $\varGamma_{ij}=\varGamma^{+}_{ij}+\varGamma^{\times}_{ij}$ while we use average $\varGamma_{ij}=(\varGamma^{+}_{ij}+\varGamma^{\times}_{ij})/2$.} They also possess the same $\ell$-scaling as in the case of monochromatic waves. As discussed in \citet{Roebber2017} for PTAs, this is to be expected. Also note that unlike in the case of monochromatic waves, the parity-violating modes $C_\ell^{TB}$ and $C_\ell^{EB}$ vanish. 

For completeness, we also include the remaining, non-Einsteinian polarisations. For the vectorial longitudinal polarisations, we have
\begin{align}
    C^{TT}_{\ell \geq 1} &= 2\pi\left(1-\frac{8}{9}\delta_{\ell1}\right)\snl[1]^2,\\
    C^{TE}_{\ell } &= \frac{2\pi}{\sqrt{\ell(\ell+1)}}\left(1-\frac{10}{9}\delta_{\ell1}\right)\snl[1]^2,\\
    C^{EE}_{\ell } &= C^{BB}_{\ell} = \frac{2\pi}{\ell(\ell+1)}\left(1-\frac{8}{9}\delta_{\ell1}\right)\snl[1]^2,\!\!
\end{align}
for the scalar transverse mode we have
\begin{align}
    C^{TT}_{\ell}&=\pi\left(\frac{\delta_{\ell1}}{9}+\delta_{\ell0}\right)\,,
    \\
    C^{TE}_{\ell}&=\frac{2\pi}{\sqrt{\ell(\ell+1)}}\frac{\delta_{\ell1}}{9}\,,
    \\
    C^{EE}_{\ell}&=\frac{4\pi}{\ell(\ell+1)}\frac{\delta_{\ell1}}{9}\,,
\end{align}
and for scalar longitudinal mode we have
\begin{align}
    C^{EE}_{\ell}&=\frac{{2\pi}}{\ell(\ell+1)}
    \left(1-\frac{8}{9}\delta_{\ell1}\right)\,.
\end{align}
For the longitudinal polarisation in the distant star limit, closed-forms of $\varGamma_{zz}(\Theta)$ and $\varGamma_{z\theta}(\Theta)$ do not exist \cite{Mihaylov2018,Lee2008}.\footnote{We believe this is why the method in \citet{Mihaylov2020,O'Beirne2018} to obtain $C^{EE}_{\ell}$ for the longitudinal polarisation fails.} For this reason, there is no $C^{TT}_{\ell}$ or $C^{TE}_{\ell}$. In the more physical scenario of stars at finite distances, they will exist but must be calculated numerically.

Having defined the angular spectra of all possible cross-correlations sourced by all polarisations, we show how these can be used to generate a realisation of the signals in the sky in Fig.~\ref{fig:gaiamap}. The maps are generated by making use of existing spin-weighted spherical harmonic routines in the \texttt{HEALPix} package \cite{2005ApJ...622..759G}. The package already includes the ability to generate spin-1 observables and we can make use of the existing visualisation tools. We generate maps of timing residuals that are correctly correlated with past values of astrometric deflections through the $TE$ correlation. The realisation only includes contributions from transverse-traceless tensor polarisations. Notice that the $TE$ contribution, which defines a correlation between the maps, is only present for the case where the time lag between observables is non-zero (see Appendix~\ref{sec:appendixA}), which would generally be the case. The signal is normalised to a stochastic background amplitude $\varOmega_\mathrm{gw}(f_0=\SI{50}{Hz})=\num{1e-8}$ which is the current upper limit for cosmological backgrounds from LIGO (see e.g. \cite{2019PhRvD.100f2001A,2019PhRvD.100f3527R}). For simplicity, we assume the signal in both timing and deflection observables is bandwidth-limited by a total integration time $T$ and fixed observation cadence of $1/\Delta t$, such that the minimum and maximum frequencies are $f_\mathrm{min}=1/T$ and $f_\mathrm{max}=1/\Delta t$ respectively. We integrate the spectral density over this frequency range to obtain the normalisation of the maps.

High resolution, full-sky realisations such as these can be useful in simulating actual observation strategies through inhomogeneous sampling and the addition of correlated noise. As a simple example, we consider the noise for an astrometric survey. We define a Fourier domain white noise amplitude $\sigma_f$ for the same bandwidth and assume the deflection field is sampled by a homogeneous distribution of $N_\star$ stars in the sky. We assume the sky is pixelated into $N_\text{pix}$ pixels of equal area and that the source count per pixel $n_\star = N_\star/N_\text{pix}$ is constant. The effective resolution per pixel is determined by the integral of the noise spectrum and can be related to an angular noise spectrum via \cite{Knox:1995dq} $N_\ell = \sigma_\text{pix}^2 \varOmega_\text{pix}$, where the variance per pixel $\sigma^2_\text{pix}$ is given by $\sigma^2_\text{reso}/n_\star$, $\sigma_\text{reso}$ defines the angular resolution of the integrated observation at each star, and $\varOmega_\text{pix}$ is the area of each sky pixel.

\begin{figure}[!t]
    \centering
    \includegraphics[width=\linewidth,height=0.85\linewidth]{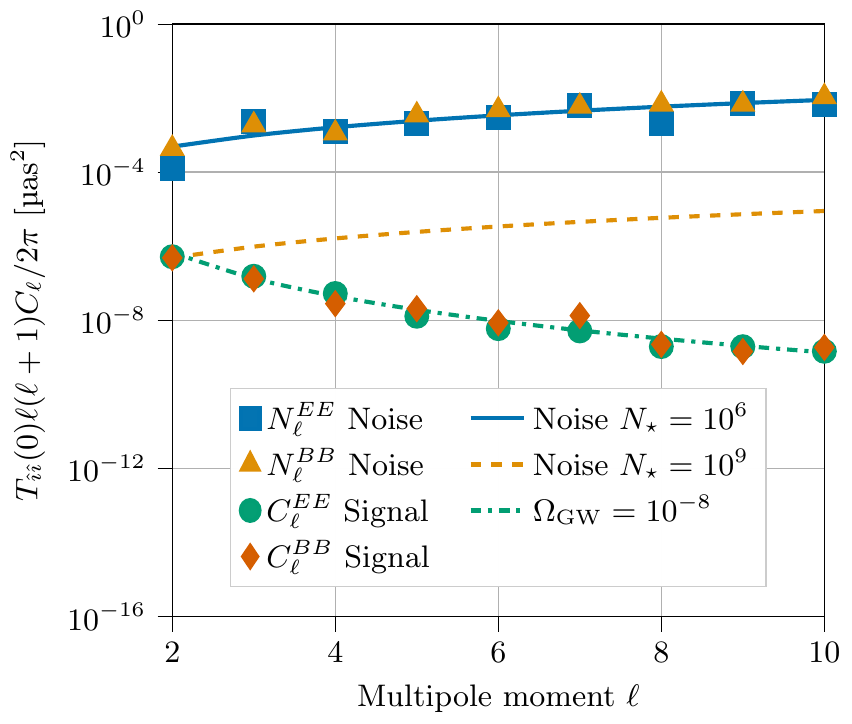}
    \caption{Angular power spectra $N_\ell=N_\ell^{EE}=N_\ell^{BB}$ of noise with angular resolution \SI{10}{\micro as} assuming $N_\star = 10^6$ (blue solid line) and $10^9$ (orange dashed line) stars, and cosmological stochastic signal with $\varOmega_\mathrm{gw}(f_0=\SI{50}{Hz})=\num{1e-8}$ and $\beta=0$ (green dash-dotted line). The physical scaling is provided by $T_{\hat\imath\hat\imath}(0)=T_{\theta\theta}(0)=T_{\phi\phi}(0)$ (see Appendix~\ref{sec:appendixA}).
    Markers represent angular spectra calculated from the multipoles used to generate the map in Fig.~\ref{fig:gaiamap}.}
    \label{fig:power}
\end{figure}

\begin{figure}[!t]
    \centering
    \includegraphics[width=\linewidth,height=0.85\linewidth]{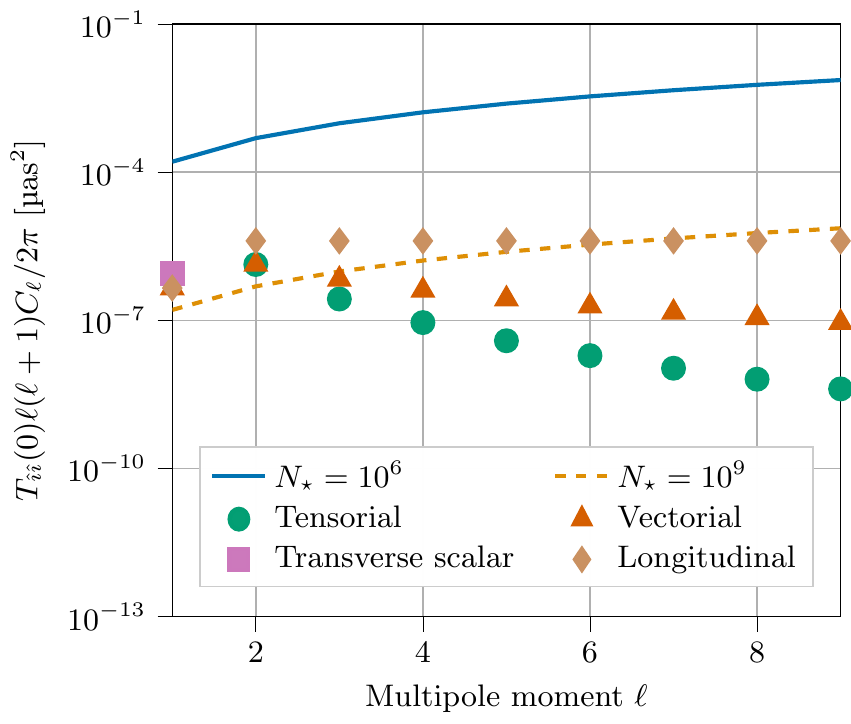}
    \caption{Angular power spectra $N_\ell^{EE}=N_\ell^{BB}$ of noise with angular resolution \SI{10}{\micro as} assuming $10^6$ (blue solid line) and $10^9$ (orange dashed line) stars. Markers represent $C_\ell^{EE}=C_\ell^{BB}$ for tensorial and vectorial polarisations and $C_\ell^{EE}$ for scalar polarisations (note that $C_\ell^{BB}=0$) of a stochastic cosmological signal with each of these polarisations assuming $\varOmega_\mathrm{gw}(f_0=\SI{50}{Hz})=\num{1e-8}$ and $\beta=0$. The physical scaling is provided by $T_{\hat\imath\hat\imath}(0)=T_{\theta\theta}(0)=T_{\phi\phi}(0)$ (see Appendix~\ref{sec:appendixA}).
    }
    \label{fig:polar}
\end{figure}

Fig.~\ref{fig:power} shows the input signal spectra for the deflection realisations along with noise spectra. We consider two headline values for the number of stars, $N_\star=10^6$ and $10^9$. The white noise amplitude is set such that the effective angular resolution for each star is \SI{10}{\micro as}. A conservative estimate for $N_\star$ lies within this range for state-of-the-art surveys, such as the results expected by the final data release for the GAIA mission \cite{gaia}. We also show the angular spectra obtained from the realisations which agree with the input values. In Fig.~\ref{fig:polar}, we show a similar set of spectra but for a selection of non-Einsteinian polarisations. We assume the same amplitude cosmological SGWB and the same noise spectra for comparison.

\section{Signal to noise ratios}\label{sec:snr}%
Signal-to-Noise Ratio (SNR) estimates for PTAs have been considered in the literature (see below). We carry out an estimate for both stochastic backgrounds and monochromatic sources for the astrometric case. We leave estimates of the cross-correlation between timing residuals and astrometry for further work.

\subsection{Stochastic backgrounds}%
We calculate the SNR statistic, $\rho$, for the astrometric components following the ``frequentist'' approach introduced for timing residuals in \citet{Moore2015}. We consider the $\rho_{ij}$ for the cross-correlation signal of $N_\star$ stars or pulsars,
\begin{equation}
    \rho_{ij}^2={8 T}\left[\int_{f_{\min }}^{f_{\max }} \mathrm{d} f \sum_{I=1}^{N_\star} \sum_{J>I}^{N_\star} \frac{|\bar\varGamma_{IJ,ij}(f)|^{2} S_{h}^{2}(f)}{P_{I}(f) P_{J}(f)}\right]\,,
\end{equation}
where the frequency bounds are, once again, set by the total integration time $T$ and observation cadence $1/\Delta t$; $P_{I}(f)$ are the noise power spectral densities for each star. The noise is assumed to be stationary in obtaining this expression for the SNR.

We next assume that the noise spectral density is white and identical for every star \cite{Thrane2013}, \begin{equation}
    P_{I}(f)=2\,\sigma_f^2 = 2 \,\Delta t\, \sigma^{2}\,,
\end{equation}
where $\sigma$ is the noise standard deviation in appropriate units
of the variable being considered (time or angle).

The overlap function in the SNR expression has a frequency dependence of
\begin{equation}
    |\bar\varGamma_{IJ,ij}(f)|^{2}=\frac{1}{(2\pi f)^{2\kappa_{ij}}}\varGamma^{2}_{ij}(\Theta_{IJ})\,,
\end{equation}
where $\kappa_{ij}$ is an exponent that depends on the correlated observables (see Eq.~\eqref{eq:nonf} for definition of $\kappa_{ij}$).
For simplicity, we assume a uniform distribution of stars in the sky. This simplifies the sampling of $\varGamma^{2}_{ij}(\Theta_{IJ})$ and in this limit, the geometric factor in the expression reduces to an overall factor 
\begin{equation*}
\chi_{ij}^2=
    \frac{2}{N^2_\star-N^{\phantom{2}}_\star}\!\sum_{I=1}^{N_\star}\!\sum_{J>I}^{N_\star} \varGamma^{2}_{ij}(\Theta_{IJ})\approx\!\frac{1}{2}\!\int_{-1}^{1} \!\! \varGamma_{ij}^{2}(\Theta)\mathrm{d}(\cos\Theta)\,,
\end{equation*}
which is the mean-square of $\varGamma_{ij}(\Theta_{IJ})$.\footnote{In our case we use a different normalisation so the numerical value of $\chi_{ij}$ will be different from $\chi^\prime$ in \citet{Moore2015}.}

In order to find the sensitivity to the characteristic strain, we have to relate it to the one-sided spectral density, $h_\text{c}(f)=\sqrt{fS_h(f)}$. The SNR in terms of the characteristic strain is then
\begin{equation}\label{eq:snr}
    \rho_{ij}^2=\frac{\chi_{ij}^2T(N^2_\star-N^{\phantom{2}}_\star)}{4\sigma^4\Delta{t}^2(2\pi)^{2\kappa_{ij}}}\left[\int_{1/T}^{1/\Delta{t}} \!\!\mathrm{d} f\; \frac{h^4_\text{c}(f)}{f^{2+2\kappa_{ij}}}\right].\!\!
\end{equation}

We set a threshold value of SNR $\rho_{ij}=3$ for detection and assume a power law
$
   h_\text{c}(f,\alpha)=A_\alpha(f/f_0)^\alpha\,.
$
Solving Eq.~\eqref{eq:snr} for $A_\alpha$ and substituting it back into the power law gives us a family of parametric curves
\begin{align}
    \displaystyle h_\text{c}(f,\alpha)&= \sigma\sqrt{\frac{\rho_{ij}\Delta t}{\chi_{ij}}}\!\left[\frac{(2\pi)^{2\kappa_{ij}}}{T(N^2_\star-N^{\phantom{2}}_\star) }\frac{\lambda_{ij}}{\left(T^{\lambda_{ij}}-\Delta{t}^{\lambda_{ij}}\right)}\right]^{\mathrlap{1/4}}\;f^{\alpha},\!\!\notag\\
    \lambda_{ij}&\equiv1+2\kappa_{ij}-4\alpha\,.
\end{align}
This represent a spectral boundary for detection of $h_\text{c}(f)$ assuming a spectral index $\alpha$ that can be observed with SNR $\rho_{ij}$. The sensitivity curve is an envelope of this family of parametric curves. This has been done for PTAs \cite{Moore2015} where $\kappa_{rr}=2$. In Fig.~\ref{fig:scurve}, we show the sensitivity curve for astrometry (where $\kappa_{\hat\imath\hat\imath}=0$) and compare it to the curve for PTAs. It is immediately obvious that, due to the different frequency dependence of the two observation methods, the sensitivity curves have very different slopes in frequency. In particular, the sensitivity for astrometric observables is shallower. There is, therefore, a potential to gain sensitivity in a frequency band that is currently gapped between PTAs and space interferometers at \SIrange{1e-6}{1e-4}{Hz}. The high-frequency limit is set by the cadence of observations $1/\Delta t$, but the flat frequency dependence means there is much to gain by increasing the speed at which the sky is surveyed. This option warrants further investigation.

\subsection{Monochromatic source}
We can also apply the estimate of the analytic sensitivity curve to the case of a monochromatic source \cite{Moore2015}. The derivation is similar to that developed for PTAs except that we generalise the source frequency dependence to include the astrometric case,
\begin{equation*}
    \tilde{h}_{I,i}(f) \approx \left(\chi \frac{h_\text{c}}{f^{\kappa_i}}  \right) \delta\left(f-f_{0}\right)\,,
\end{equation*}
where again, $\kappa_r=1$ for PTAs and $\kappa_{\hat\imath}=0$ for astrometry. The sky-averaged geometric factor is 
\begin{equation}
    \chi^2=\frac{1}{4\pi}\int_{S^2} \!\ud\varOmega_{\bhn}\sum_P|R^{P}_{i}(f_0,\bhk_0,\bhn)|^2=\frac{1}{3}\,,
\end{equation}
and it is identical for both astrometry and PTA. 
Similarly, we can write the SNR expression
\begin{equation}\label{eq:SNRmono}
    \varrho_{ij}^{2}=\frac{\chi^{4} h_{c}^{4}(N^2_\star-N^{\phantom{2}}_\star)}{\sigma^4\Delta{t}^2T} \left[\int_{1 / T}^{1 / \Delta t} \!\!\mathrm{d} f\; \frac{\delta_{T}^{4}\left(f-f_{0}\right)}{f^{2\kappa_{ij}}}\right],
\end{equation}
which can be inverted to find a sensitivity curve $h_\text{c}(f)$. 

\begin{figure}[!t]
    \centering
    \includegraphics[width=\linewidth,height=0.85\linewidth]{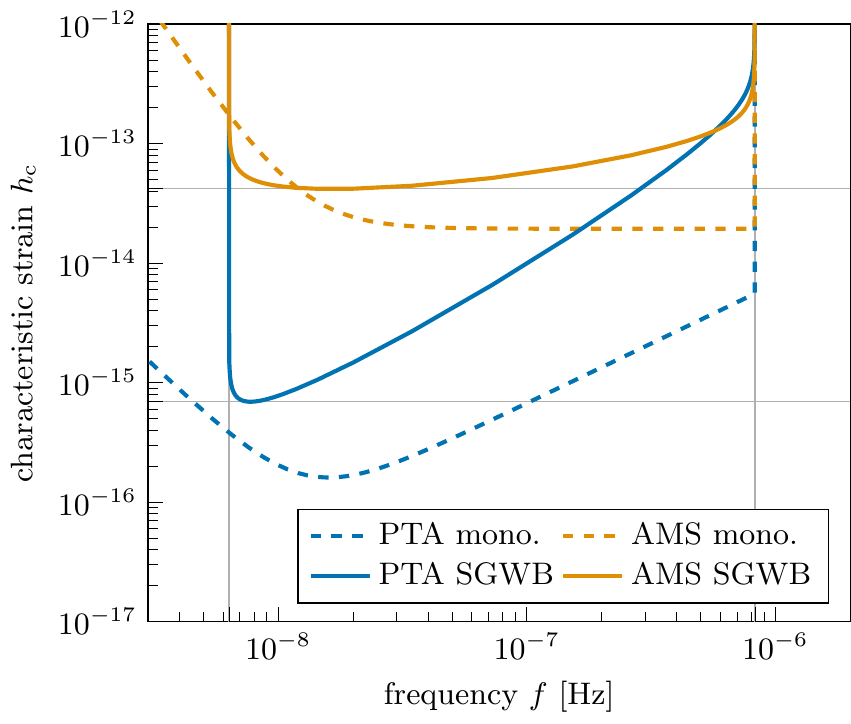}
    \caption{Frequentist analytic sensitivity curves for an astrometric survey (AMS) and pulsar timing array (PTA). All curves assume observation of 5 years measured fortnightly. The PTA sensitivity curve is included for reference and is assumed to be made of 36 pulsars with the rms error in the timing-residuals of \SI{100}{\nano\second} \cite{Moore2015}, whereas the astrometry curve assumes $10^6$ stars with measurement noise of \SI{10}{\micro as}. }
    \label{fig:scurve}
\end{figure}

As mentioned in 
\cite{Moore2015}, this curve will not account for the loss of sensitivity due to fitting out the quadratic timing/astrometric model. To account for this, we can use the same approximation as \cite{Moore2015} which allows us to change the integral in Eq.~\eqref{eq:SNRmono} from frequency to time domain. We then look at two limits; the high frequency limit $(f t \gg 1)$, which we use to find $h^\text{HIGH}_\text{c}(f)$ and the low frequency limit $(f t \ll 1)$ in which we expand the integrand (Eq.~(15) in \citet{Moore2015}) and take the contribution of order $\mathcal{O}(f^3 t^3)$ to find $h^\text{LOW}_\text{c}(f)$. We combine these two which are equivalent to their Eqs.~(14)~and~(16) to get 
\begin{align}
    \begin{split}
        \!\!h_\text{c}(f)&\approx h^\text{LOW}_\text{c}(f)+h^\text{HIGH}_\text{c}(f)\\
        &\approx \frac{\sigma}{\chi} \sqrt{\frac{\varrho_{ij}\Delta t}{T}}\!\left[\frac{16f^{2\kappa_{ij}}}{3 (N^2_\star-N^{\phantom{2}}_\star)}\right]^{1/4}\!\!\left(1+\frac{f_\text{p}^{3}}{f^{3}}\right),\!\!
    \end{split}
\end{align}
where $f_\text{p}$ is frequency at which $h^\text{LOW}_\text{c}(f_\text{p})=h^\text{HIGH}_\text{c}(f_\text{p})$ and it is also chosen to be $2/T$. We show the monochromatic sensitivity curves in Fig.~\ref{fig:scurve} alongside those for an SGWB. The frequency dependence for astrometric observations, in this case, is flat and this increases the advantage of astrometric observations even further. Our baseline assumption of $N_\star=10^6$ at \SI{10}{\micro as} is a conservative one. An ambitious goal of $N_\star\sim 10^9$ at a similar resolution at a sampling rate of \SI{1e-5}{Hz} would offer an interesting level of sensitivity in a frequency band that is complementary to other detection methods.

\section{Discussion}\label{sec:disc}%
We have introduced a polarisation-like complex spin-$s$ field description of ``astrochronometric'' observables on the sphere. This setup enables the analysis of the sphere using the spin-weighted harmonics formalism in analogy with the polarisation of the CMB. This formalism can be used to derive compact forms of the harmonic cross-spectra of observables sourced by any polarisation components of GWs. The formalism also allows a simplified relationship between the angular power spectra and coordinate domain correlation functions as shown in Eqs.~\eqref{eq:Czz}--\eqref{eq:Cnn}. These relationships have proven to be very useful in the analysis of CMB observations which necessitate robust estimation of correlations in polarisation patterns in both coordinate and harmonic domains.

The introduction of a spin-$s$ description enables us to easily create realisations of the sky in both timing and deflection observables. This will be of use in assessing the feasibility of observational strategies and the development of robust estimation tools for future data sets. This application relies on a mature infrastructure developed over several decades for analysis, simulation, and visualisation of polarised CMB observations. 

A key advantage of our formalism is that it makes the connection between the \textit{spin} of GW polarisation and the nature of the resulting anisotropies explicit. We see directly how different Einsteinian and non-Einsteinian polarisations source $\ell\leq 2$ differently and how vectorial polarisations induce specific correlations in the observables. If astrochronometric observations were to become accurate enough, the search for the tell-tale presence of GW-induced dipole components might provide constraints on departures from GR. Challenges remain, however. The presence of a kinematic dipole due to the observer's motion relative to the cosmological rest frame, along with higher multipoles due to acceleration, may prove to be an insurmountable obstacle. We leave for future work a calculation of SNR for individual multipoles to constrain individual polarisations.

We have also presented an estimate of signal-to-noise ratio statistics for astrometry. Our results show that astrometric observations, and their correlations with timing observations, may provide a complementary window in frequency to a PTA--style analysis. The possibility here is that a fast scan strategy at current levels of angular resolution may provide interesting constraints at frequencies \SIrange{e-6}{e-5}{Hz} that are between the PTA and LISA windows.

\begin{acknowledgments}
    CRC acknowledges support under a UKRI Consolidated Grant ST/T000791/1. SG is supported by a UKRI Studentship under grant ST/T506151/1.
\end{acknowledgments}

\appendix%
\section{Realisations of the sky}\label{sec:appendixA}%
We summarise how the lagged cross-correlations between deflection and timing residual signals are related to the frequency domain spectra of the underlying gravitational wave backgrounds. 

\begin{figure}[t]
    \centering
    \includegraphics[width=\linewidth,height=0.67\linewidth]{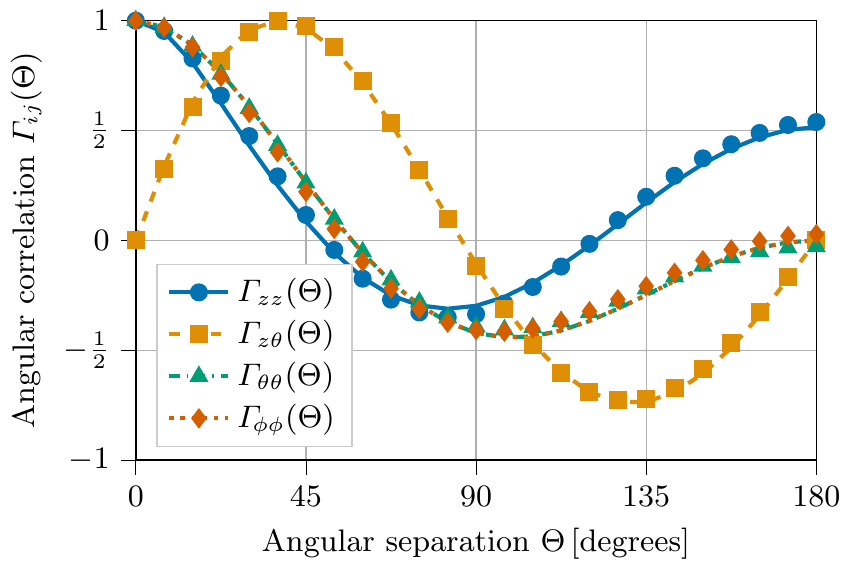}
    \caption{Tensorial normalised angular correlations $\varGamma_{zz}$ (blue solid line), $\varGamma_{z\theta}$ (orange dashed line) and $\varGamma_{\theta\theta}=\varGamma_{\phi\phi}$ (green dash-dotted and red dotted lines). Markers represent the correlations sampled in the realisations shown in  Fig.~\ref{fig:gaiamap}. 
    }
    \label{fig:angular}
\end{figure}

\begin{figure}[t]
    \centering
    \includegraphics[width=\linewidth,height=0.67\linewidth]{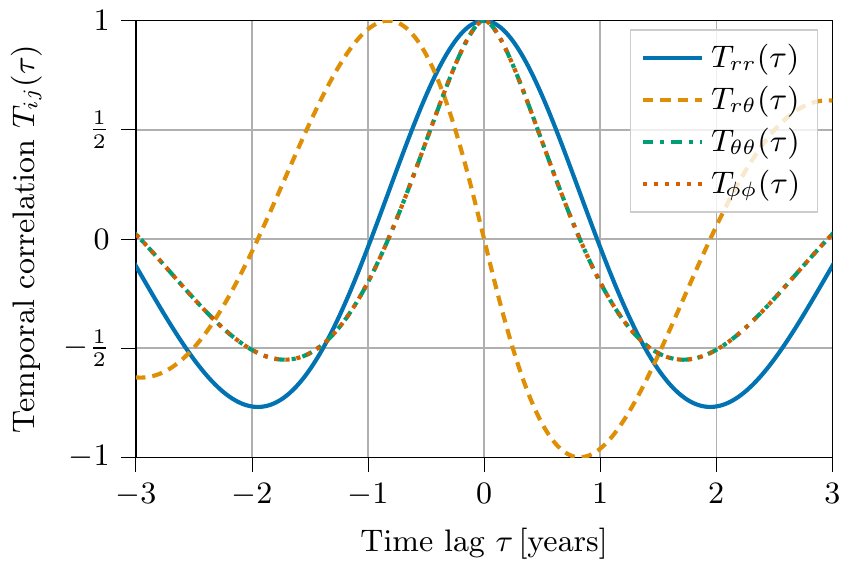}
    \caption{Normalised temporal correlations with $\beta=0$, $T=\SI{5}{years}$ and $\Delta{t}=\SI{14}{days}$ for residual--residual $T_{rr}$ (blue solid line), residual--deflection $T_{r\theta}$ (orange dashed line) and deflection--deflection $T_{\theta\theta}=T_{\phi\phi}$ (green dash-dotted and red dotted lines).}
    \label{fig:temporal}
\end{figure}

Let us assume we observe a set of $N_\star$ stars. For every star $I$ at each epoch $t$ we have a timing residual, $r_{I}(t)$, and astrometric deflection, $\delta n_{I,{\hat\imath}}(t)$. We combine these to a vector $h_{I,i}(t)=(r_{I}(t),\delta{n}_{I,{\hat\imath}}(t))^\intercal$ and cross-correlate 
\begin{align}
    \left(h_{I,i} \star h_{J,j}\right)(\tau)&=\left\langle h_{I,i}(t) h_{J,j}(t+\tau)\right\rangle\notag\\
    &=T_{ij}(\tau) \varGamma_{ij}(\Theta_{I J}),
\end{align}
where in the last line we have separated the correlation into angular $\varGamma_{ij}\left(\Theta_{I J}\right)$ and temporal $T_{ij}(\tau)$ parts. The correlation structure that is normally considered in angular separation is also reflected in the temporal correlations for different lags $\tau$. We show both correlations in Figs.~\ref{fig:angular} and~\ref{fig:temporal}. Notice how the cross-correlation of timing residuals with deflections is odd with respect to the lag and that, in general, the amplitude of the cross-correlation will depend on the lag along with the total integration time which sets the effective frequency high-pass (see below).

Thanks to the property of cross-correlation and Fourier transform, $\mathcal{F}\{f\star g\}=\mathcal{F}\{f\}\mathcal{F}\{g\}$, we can find what the temporal correlations are from correlations in Fourier space that are
\begin{align}
    \langle \tilde{h}^{\phantom{\ast}}_{I,i}(f) \tilde{h}^\ast_{J,j}(f^\prime)\rangle&=\frac{1}{2}\delta\left(f-f^{\prime}\right)S_{h}(f)\bar\varGamma_{IJ,ij}(f)  \,, \\
    \bar\varGamma_{IJ,ij}(f)&=\frac{(-\ui)^{\kappa_i}(\ui)^{\kappa_j}}{(2\pi f)^{\kappa_{ij}}}\varGamma_{ij}(\Theta_{IJ})  \,,
\end{align}
where $\kappa_i=\delta_{ir}$ and $\kappa_{ij}=\kappa_i+\kappa_j$ are exponents coming from the the fact that the timing residual is an integrated redshift.
We can use $T_{ij}(\tau)$ as a physical scaling of our $C_\ell$, which we can use to plot maps of realisations of $r_{I}(t)$ and $\delta n_{I{\hat\imath}}(t-\tau)$ in the sky (see Fig~\ref{fig:gaiamap}). This scaling is
\begin{alignat}{2}
    T_{ij}(\tau)&=\operatorname{Re}\left\{\int_{1/T}^{1/\Delta{t}} \!\mathrm{d} f \;{e^{-2\pi \ui f \tau}}S_{h}(f)\frac{(-\ui)^{\kappa_i}(\ui)^{\kappa_j}}{(2\pi f)^{\kappa_{ij}}} \right\}\\\notag
    &=\phantom{-}\frac{6 H_{0}^{2}}{(2 \pi)^{2+\kappa
    _{ij}}} \frac{\varOmega_\mathrm{gw}\left(f_{0}\right)}{f_{0}^{\beta}}\notag\\ &\times\operatorname{Re}\left\{t^{\kappa_{ij}-\beta+2}\operatorname{E}_{\kappa_{ij}-\beta+3}{\left(2\pi\ui\frac{\tau}{t}\right)}(-\ui)^{\kappa_i}(\ui)^{\kappa_j}\right\}\Big|_{t=T}^{\Delta{t}},\notag\!\! 
\end{alignat}
where we have introduced the generalised exponential integral
\begin{equation}
    \operatorname{E}_{n}(z)=\int_{1}^{\infty} \frac{\ue^{-z \nu}}{\nu^{n}} \mathrm{d}\nu\,.
\end{equation}

\section{Polarisation basis}\label{sec:pol}%
For completeness, we include a summary of the polarisation conventions used as there are a number of different conventions adopted in the literature.

Given any orthonormal coordinate system $x$, $y$, $z$ we can find a frame associated with spherical coordinates at point $\bm{r}=(r,\vartheta,\varphi)^\intercal$ as follows
\begin{align}
    \hat{\bm{e}}^{r}({\bm{r}})&=\hat{\bm{r}},&\!\!
    \hat{\bm{e}}^{\varphi}({\bm{r}})&=\frac{\hat{\bm{z}}\times\hat{\bm{r}}}{|\hat{\bm{z}}\times\hat{\bm{r}}|},&\!\!
    \hat{\bm{e}}^{\vartheta}({\bm{r}})&= \hat{\bm{e}}^{\varphi}({\bm{r}})\times\hat{\bm{r}}.
\end{align}
This vector frame can be used to define a complete set of six polarisation basis tensors for any general polarisation.
We can sort them into tensorial modes
\begin{align}
    e_{ab}^{+}(\bhk)&=\hat{e}_{a}^{\vartheta} \hat{e}_{b}^{\vartheta}-\hat{e}_{a}^{\varphi} \hat{e}_{b}^{\varphi}, &
    e_{ab}^{\times}(\bhk)&=\hat{e}_{a}^{\vartheta} \hat{e}_{b}^{\varphi}+\hat{e}_{a}^{\varphi} \hat{e}_{b}^{\theta}, \\
    \intertext{vectorial modes}
    e_{ab}^{X}(\bhk)&=\hat{e}_{a}^{\vartheta} \hat{e}_{b}^{r}+\hat{e}_{a}^{r} \hat{e}_{b}^{\vartheta}, &
    e_{ab}^{Y}(\bhk)&=\hat{e}_{a}^{\varphi} \hat{e}_{b}^{r}+\hat{e}_{a}^{r} \hat{e}_{b}^{\varphi}, \\
    \intertext{and scalar modes}
    e_{ab}^{S}(\bhk)&=\hat{e}_{a}^{\vartheta} \hat{e}_{b}^{\vartheta}+\hat{e}_{a}^{\varphi} \hat{e}_{b}^{\varphi},&
    e_{ab}^{L}(\bhk)&=\sqrt{2} \hat{e}_{a}^{r} \hat{e}_{b}^{r} .
\end{align}

\section{Frequency-independent overlap reduction functions}\label{sec:gamma}%
In Section~\ref{sec:SGWB}, we calculate the angular power spectra of the stochastic gravitational background. For that, we need overlap reduction functions
\begin{align}    \label{eq:nofg}
    \varGamma_{ij}(\mu)&=\frac{1}{4 \pi g}\int_{S^2} \!\ud\varOmega_{\bhk}
    \sum\limits_{P}
    R_{i}^{P}(\bhn,{\bhk})
    {R_{j}^{P}}^\ast\!(\bhn^{\mathrlap{\prime}},{\bhk})\,,
\end{align}
which we present below in terms of $\mu=\bhn\cdot\bhn^\prime=\cos{\Theta}$, ignoring co-located pulsar terms. These functions are related to those in \citet{Mihaylov2018} by a factor of $4 \pi g$.
For an isotropic background, the symmetries will ensure that there will only be four overlap reduction functions (see e.g. \cite{Mihaylov2018,O'Beirne2018,Book2011}) in the matrix,\footnote{Note that in \citet{O'Beirne2018}, they use different notation $\sigma=g\varGamma_{\theta \theta}$ and $\alpha=g\varGamma_{\phi \phi}$. In \citet{Book2011}, they use similar notation but their $\sigma$ has the opposite sign.}
\begin{equation}
    \varGamma(\mu)=\left(\begin{array}{ccc}
    \varGamma_{z z}(\mu) & \varGamma_{z \theta}(\mu) & 0 \\
    \varGamma_{z \theta}(\mu) & \varGamma_{\theta \theta}(\mu) & 0 \\
    0 & 0 & \varGamma_{\phi \phi}(\mu)
    \end{array}\right)\,.
\end{equation}
These can be translated to a harmonic space representation using Eqs.~\eqref{eq:Czz}--\eqref{eq:Cnn}.
Tables~\ref{tab:tensor}--\ref{tab:long} list the first few coefficients of the harmonic space representation.
\subsection{Tensorial polarisations}%
For tensorial polarisations, there are only three independent functions, i.e. $\varGamma_{\theta \theta}(\mu)=\varGamma_{\phi \phi}(\mu)=\varGamma_{\hat\imath \hat\imath}(\mu)$
\begin{align}
    \varGamma_{zz}(\mu)&=\frac{1}{8}+\frac{\mu}{24}+\frac{1-\mu}{4} \ln{ \left(\frac{1-\mu}{2}\right)}\,,\\
    \varGamma_{\phantom{z}\mathclap{z\theta}\phantom{z}}(\mu)&=
    \frac{\sqrt{1-\mu^2}}{6}+\frac{1}{4}\frac{(1-\mu)^2}{\sqrt{1-\mu^2}} \ln{\left(\frac{1-\mu}{2}\right)}\,,\\
    \varGamma_{\phantom{z}\mathclap{\hat\imath\hat\imath}\phantom{z}}(\mu)&=-\frac{5}{24}+\frac{7 \mu}{24}- \frac{1}{4}\frac{(1-\mu)^{2}}{1+\mu} \ln{\left(\frac{1-\mu}{2}\right)}\,.
\end{align}
The redshift correlation comes from \citet{Hellings} and the remaining functions from \citet{Mihaylov2018}.

\begin{table}[!ht]
    \centering
    \caption{First six $C_\ell$ of tensorial polarisations $C_{\ell}^{\mathrm{BB}}=C_{\ell}^{EE}$.}\label{tab:tensor}
    \begin{ruledtabular}
        \begin{tabular}{c|cccccc} 
            $\ell$  & 2 & 3 & 4 & 5 & 6 & 7  \\
            \hline$C_{\ell}^{T T}$ & $\dfrac{\mathstrut \pi}{\mathstrut 12}$ & $\dfrac{\pi}{60}$ & $\dfrac{\pi}{180}$ & $\dfrac{\pi}{420}$ & $\dfrac{\pi}{840}$ & $\dfrac{\pi}{1512}$  \\
            $C_{\ell}^{T E}$ & $\dfrac{\mathstrut \pi}{\mathstrut 6 \sqrt{6}}$ & $\dfrac{\pi}{60 \sqrt{3}}$ & $\dfrac{\pi}{180 \sqrt{5}}$ & $\dfrac{\pi}{210 \sqrt{30}}$ & $\dfrac{\pi}{420 \sqrt{42}}$ & $\dfrac{\pi}{1512 \sqrt{14}}$  \\
            $C_{\ell}^{EE}$  & $\dfrac{\mathstrut \pi}{\mathstrut 18}$ & $\dfrac{\pi}{180}$ & $\dfrac{\pi}{900}$ & $\dfrac{\pi}{3150}$ & $\dfrac{\pi}{8820}$ & $\dfrac{\pi}{21168}$  \\
        \end{tabular}
    \end{ruledtabular}
\end{table}

\subsection{Vectorial polarisations}%
For vectorial polarisations, there are only three independent functions, i.e. $\varGamma_{\theta \theta}(\mu)=\varGamma_{\phi \phi}(\mu)=\varGamma_{\hat\imath \hat\imath}(\mu)$
\begin{align}
    \varGamma_{zz}(\mu)&=-\frac{1}{2}-\frac{2\mu}{3}-\frac{1}{2} \ln \left(\frac{1-\mu}{2}\right)\,,\\
    \varGamma_{\phantom{z}\mathclap{z\theta}\phantom{z}}(\mu)&=
    -\frac{5}{12}\sqrt{1-\mu^{2}}-\frac{1}{2}\sqrt{\frac{1-\mu}{1+\mu}} \ln \left(\frac{1-\mu}{2}\right)\,,\\
    \varGamma_{\phantom{z}\mathclap{\hat\imath\hat\imath}\phantom{z}}(\mu)&=\frac{1}{3}-\frac{\mu}{6}+\frac{1}{2}\frac{1-\mu}{1+\mu} \ln \left(\frac{1-\mu}{2}\right)\,.
\end{align}
All functions are from \citet{Mihaylov2018} with an appropriate scaling.

\begin{table}[!ht]
    \centering
    \caption{First six $C_\ell$ of vectorial polarisations $C_{\ell}^{\mathrm{BB}}=C_{\ell}^{EE}$.}\label{tab:vector}
    \begin{ruledtabular}
        \begin{tabular}{c|cccccc} 
            $\ell$ & 1 & 2 & 3 & 4 & 5 & 6 \\
            \hline$C_{\ell}^{TT}$  & $\dfrac{\mathstrut\pi}{\mathstrut9}$ & $\dfrac{\pi}{3}$ & $\dfrac{\pi}{6}$ & $\dfrac{\pi}{10}$ & $\dfrac{\pi}{15}$ & $\dfrac{\pi}{21}$ \\
            $C_{\ell}^{TE}$  & $-\dfrac{\mathstrut\pi}{\mathstrut9 \sqrt{2}}$ & $\dfrac{\pi}{3 \sqrt{6}}$ & $\dfrac{\pi}{12 \sqrt{3}}$ & $\dfrac{\pi}{20 \sqrt{5}}$ & $\dfrac{\pi}{15 \sqrt{30}}$ & $\dfrac{\pi}{21 \sqrt{42}}$ \\
            $C_{\ell}^{EE}$  & $\dfrac{\mathstrut\pi}{\mathstrut18}$ & $\dfrac{\pi}{18}$ & $\dfrac{\pi}{72}$ & $\dfrac{\pi}{200}$ & $\dfrac{\pi}{450}$ & $\dfrac{\pi}{882}$ \\
    \end{tabular}
    \end{ruledtabular}
\end{table}

\subsection{Scalar transverse polarisation}%
For scalar polarisations, $\varGamma_{\theta \theta}(\mu)\neq\varGamma_{\phi \phi}(\mu)$ so we will have four unique functions
\begin{align}
    \varGamma_{zz}(\mu)&= \frac{1}{4}+\frac{\mu}{12}\,,\\
    \varGamma_{\phantom{z}\mathclap{z\theta}\phantom{z}}(\mu)&=
    \frac{1}{12} \sqrt{1-\mu^{2}}\,,\\
    \varGamma_{\phantom{z}\mathclap{\theta\theta}\phantom{z}}(\mu)&=\mu\varGamma_{\phantom{z}\mathclap{\phi\phi}\phantom{z}}(\mu)=\frac{\mu}{12} \,,
\end{align}
where the $\varGamma_{zz}$ is from \citet{O'Beirne2018}. 

\begin{table}[!ht]
    \centering
    \caption{First six $C_\ell$ of scalar transverse polarisation.}\label{tab:scal}
    \begin{ruledtabular}
        \begin{tabular}{c|cccccc}
            $\ell$&0 & 1 & 2 & 3 & 4 & 5\\
            \hline
            $C_{\ell}^{T T}$& ${\pi}$& $\dfrac{\mathstrut\pi}{\mathstrut9}$&0&0&0&0\\
            $C_{\ell}^{T E}$&0& $\dfrac{\mathstrut \pi\sqrt{2}}{\mathstrut9}$&0&0&0&0 \\
            $C_{\ell}^{E E}$&0& $\dfrac{\mathstrut2 \pi}{\mathstrut9}$&0&0&0&0
        \end{tabular}
    \end{ruledtabular}
\end{table}

\subsection{Scalar longitudinal polarisation}%
In the distant star limit, the redshift and redshift--deflection correlations are undefined \cite{Mihaylov2018,Lee2008}. Nonetheless, the remaining functions are
\begin{align}
    \varGamma_{\phantom{z}\mathclap{\theta\theta}\phantom{z}}(\mu)&=-\frac{1}{2}-\frac{\mu}{3} -\frac{1}{2}\frac{1}{(1+\mu)} \ln \left(\frac{1-\mu}{2}\right)\,,\\
    \varGamma_{\phantom{z}\mathclap{\phi\phi}\phantom{z}}(\mu)&=-\frac{1}{3}-\frac{1}{2}\frac{1}{(1+\mu)} \ln \left(\frac{1-\mu}{2}\right) \,.
\end{align}

\begin{table}[!h]
    \centering
    \caption{First six $C_\ell$ of scalar longitudinal polarisation.}\label{tab:long}
    \begin{ruledtabular}
        \begin{tabular}{c|cccccc} 
            $\ell$ & 1 & 2 & 3 & 4 & 5 & 6 \\
            \hline$C_{\ell}^{EE}$ & $\dfrac{\mathstrut\pi}{\mathstrut9}$ & $\dfrac{\pi}{3}$ & $\dfrac{\pi}{6}$ & $\dfrac{\pi}{10}$ & $\dfrac{\pi}{15}$ & $\dfrac{\pi}{21}$
        \end{tabular}
    \end{ruledtabular}
\end{table}

\hbadness 10000\relax \bibliography{main}
\end{document}